\documentclass[journal,a4paper]{IEEEtran}
\usepackage[utf8]{inputenc} 
\usepackage[T1]{fontenc}
\usepackage{url}              % provides \url{...}
\usepackage{cite}             % improves presentation of citations

\usepackage[cmex10]{amsmath}  % Use the [cmex10] option to ensure complicance
\usepackage{mleftright}       % fix to wrong spacing of \left-,
\mleftright                   % \middle- \right-commands 

\usepackage{graphicx}         % provides \includegraphics{...} to
\usepackage{booktabs}         % fixes poor spacing in tables and
\usepackage{amsmath,amsfonts,bm}
\usepackage{mathrsfs}
\usepackage{algorithmic}
\usepackage{algorithm}
\usepackage{array}
\usepackage[caption=false,font=normalsize,labelfont=sf,textfont=sf]{subfig}
\usepackage{textcomp}
\usepackage{stfloats}
\usepackage{url}
\usepackage{verbatim}
\usepackage{graphicx}
\usepackage{cite}
\usepackage{tikz}
\usetikzlibrary{shapes,arrows}
\usepackage{xcolor}

\def \a {{\bf a}}
\def \b {{\bf b}}
\def \c {{\bf c}}

\def \z {{\bf z}}

\newtheorem{theorem}{\textbf{Theorem}}[section]
\newtheorem{lemma}{\textbf{Lemma}}[section]

\newtheorem{definition}{\textbf{Definition}}[section]
\newtheorem{proposition}{\textbf{Proposition}}[section]
\newtheorem{remark}{\textbf{Remark}}[section]
\newtheorem{example}{\textbf{Example}}[section]

\begin{document}

\title{ Bounds on Size of Homopolymer Free Codes } 

\author{
 \IEEEauthorblockN{  Krishna Gopal Benerjee, and Adrish Banerjee} \thanks{The work is accepted in IEEE International Symposium on Information Theory (ISIT) 2023.} \\
   \IEEEauthorblockA{ Department of Electrical Engineering, Indian Institute of Technology Kanpur, India \\
					  Email: \{kgopal, adrish\}@iitk.ac.in           
                     }
 }

\maketitle

\begin{abstract}
For any given alphabet of size $q$, a Homopolymer Free code (HF code) refers to an $(n, M, d)_q$ code of length $n$, size $M$ and minimum Hamming distance $d$, where all the codewords are homopolymer free sequences.
For any given alphabet, this work provides upper and lower bounds on the maximum size of any HF code using Sphere Packing bound and Gilbert-Varshamov bound. 
Further, upper and lower bounds on the maximum size of HF codes for various HF code families are calculated. 
Also, as a specific case, upper and lower bounds are obtained on the maximum size of homopolymer free DNA codes.
\end{abstract}

\section{Introduction}\label{Sec: Introduction}
DeoxyriboNucleic Acid (DNA) is employed in several DNA computing applications, including DNA-based data storage. 
DNA-based data storage has a high storage density, capacity, and endurance \cite{Church1628}.
Errors like insertion, deletion, and flip are typically seen when sequencing and synthesizing DNA. 
One of the causes of these errors is the DNA with numerous consecutive repeats of a nucleotide (called a homopolymer) with a high repetition factor \cite{8444440,goldman2013towards,8827908}.
Therefore, DNA codes that avoid homopolymers are preferred \cite{goldman2013towards,Benerjee2021,8827908,9032172,1228122,9174438,9380305,9214530,9834744,9965847,8695057,8772194,9611362}.
In fact, in \cite{goldman2013towards,Benerjee2021}, DNA codes are obtained that are free from homopolymers.
Further, bounds on various DNA codes with additional properties such as Reverse constraint, Reverse-Complement constraint, $GC$-content constraint, and avoiding homopolymers are discussed in \cite{King2003,9929400,GABORIT200599,doi:10.1089/10665270152530818,math10050845,gaborit2005linear,4595399,Bennenni2019}.
Lower bounds on the maximal size of DNA code with $GC$-content constraint is given in \cite{8424143}, where all DNA codewords are independent of homopolymers. 
In \cite{King2003,9929400}, authors have investigated bounds on the size of DNA codes with the $GC$-content constraint.
It is interesting to study the upper and lower bounds of the size of homopolymer free codes in general.

For any alphabet of size $q$, an $(n,M,d)_q$ \textit{code} is a set of $n$-length sequences and size $M$ such that the minimum distance $d$ = $\min\limits_{\a,\b\in\mathcal{C};\a\neq\b}d(\a,\b)$, where $d(\a,\b)$ is the Hamming distance between sequences $\a$ and $\b$.
In particular, any code defined over the alphabet $\{A, C, G, T\}$ is called a \textit{DNA code}.
Also, for any given integers $n$, $q$ and $d$, if $A_q(n,d)$ = $\max\limits_{\mathcal{C}}M$, where $\mathcal{C}$ is an $(n,M,d)_q$ code, then
\begin{itemize}
    \item Sphere Packing (SP) bound: 
    \begin{equation}
        A_q(n,d)\leq \frac{V_q(n)}{\sum_{r=0}^{\lfloor (d-1)/2\rfloor}V_q(n,r)},
        \label{SP bound Gen}
    \end{equation}
    \item Gilbert-Varshamov (GV) bound:
    \begin{equation}
        A_q(n,d)\geq \frac{V_q(n)}{\sum_{r=0}^{d-1}V_q(n,r)},
        \label{GV bound Gen}
    \end{equation}
\end{itemize}
where $V_q(n)$ is the full size of the space and $V_q(n,r)$ is the number of sequences that differ by Hamming distance $r$ with a given codeword. Also note that, for the given code and $r$, it is assumed that the error ball size $V_q(n,r)$ is uniform for each codeword in inequality (\ref{SP bound Gen}) and inequality (\ref{GV bound Gen}).
For more details on such bounds, refer to \cite{10.1216/RMJ-1975-5-2-199,1023642,4418465,Zhang:2010:1546-1955:2072,7891597,9328218,9515944,612999,1523339,866628,8937735,5452190,9465135,7467537,623158,7061480,e24101473}.
In general, for a code without any additional property, $V_q(n)=q^n$ and $V_q(n,r) = \binom{n}{r}(q-1)^r$. 
Also, the homopolymer free property for the sequence is defined as follows.
A sequence over the $q$-size alphabet is called a homopolymer free sequence if symbols at consecutive locations are not the same. 
For example, the $5$-length sequence $\alpha_1\alpha_3\alpha_1\alpha_2\alpha_3$ is a homopolymer free sequence defined over the alphabet $\{\alpha_1,\alpha_2,\alpha_3\}$.

The main contributions of this paper are as follows: 
\begin{itemize}
\item In this paper, upper and lower bounds on the size of HF codes over an $q$-size alphabet are obtained using Sphere Packing and Gilbert-Varshamov bounds.
\item Then, as a special case, upper and lower bounds on the maximum size of HF DNA codes are reported.
\end{itemize}
\textit{Organisation}: Preliminary background and notations are described in Section \ref{Sec: Preliminary}. 
For homopolymer free codes, bounds and results are obtained in Section \ref{Sec: HF Codes}.
Particularly, for homopolymer free DNA codes, bounds and results are described in Section \ref{Sec: HF DNA Codes}.
Also, comparisons for homopolymer free DNA codes are given in Section \ref{Sec: Discussion}.

\section{Preliminaries}\label{Sec: Preliminary}
The notations and definitions used in this paper have been discussed in this section. 

For any positive integer $q$, consider an alphabet $\mathcal{A}_q$ = $\{\alpha_1,\alpha_2,\ldots,\alpha_q\}$ of size $q$.
Now, we define the HF sequence and HF codes in Definition \ref{HF code def.} as follows.
\begin{definition}
A sequence $\z$ = $z_1z_2\ldots z_n$ of length $n$ over the alphabet $\mathcal{A}_q$ is called homopolymer free sequence (HF sequence) if $z_i\neq z_{i+1}$ for $i$ = $1,2,\ldots,n-1$.
An $(n, M, d)_q$ code $\mathcal{C}$ is called a homopolymer free code (HF code) if each codeword of the code is an HF sequence.
\label{HF code def.}
\end{definition}
\begin{example}
For an alphabet $\mathcal{A}_3$ = $\{\alpha_1,\alpha_2,\alpha_3\}$, consider a code $\mathcal{C}$ = $\left\{\alpha_1\alpha_2\alpha_1\alpha_3, \alpha_2\alpha_3\alpha_2\alpha_3, \alpha_2\alpha_1\alpha_3\alpha_1\right\}$ defined over the alphabet $\mathcal{A}_3$, where all the codewords are HF sequence.
Hence, the code $\mathcal{C}$ is an $(4,3,3)_3$ HF code.
\end{example}

For any integers $n$ and $r$, consider a set, $\mathcal{C}_{q,n}$, of all HF sequences each of length $n$ over $\mathcal{A}_q$.
So, a homopolymer free sphere (HF sphere) with the centre $\a$ and the radius $r$ is the set $H_r(\a) \overset{\Delta}{=} \{\z:d(\a,\z)=r\mbox{ and } \z\in\mathcal{C}_{q,n}\}$.
Also, the size of the HF sphere $H_r(\a)$ is referred by $|H_r(\a)|$.
\begin{example}
For two sequences $\alpha_1\alpha_2\alpha_3$ and $\alpha_1\alpha_2\alpha_1$ of length three over the alphabet $\mathcal{A}_5$ = $\{\alpha_1, \alpha_2, \alpha_3, \alpha_4, \alpha_5\}$, 
\begin{itemize}
    \item HF sphere is $H_1(\alpha_1\alpha_2\alpha_3)$ = $\left\{\alpha_3\alpha_2\alpha_3,\alpha_4\alpha_2\alpha_3,\alpha_5\alpha_2\alpha_3\right.$, $\left.\alpha_1\alpha_4\alpha_3,\alpha_1\alpha_5\alpha_3,\alpha_1\alpha_2\alpha_1,\alpha_1\alpha_2\alpha_4,\alpha_1\alpha_2\alpha_5\right\}$, where the size of the HF sphere is $|H_1(\alpha_1\alpha_2\alpha_3)|$ = $8$, and
    \item HF sphere $H_1(\alpha_1\alpha_2\alpha_1)$ = $\left\{\alpha_3\alpha_2\alpha_1,\alpha_4\alpha_2\alpha_1,\alpha_5\alpha_2\alpha_1,\right.$  $\left.\alpha_1\alpha_3\alpha_1,\alpha_1\alpha_4\alpha_1,\alpha_1\alpha_5\alpha_1,\alpha_1\alpha_2\alpha_3,\alpha_1\alpha_2\alpha_4,\alpha_1\alpha_2\alpha_5\right\}$, where the size of the HF sphere is $|H_1(\alpha_1\alpha_2\alpha_1)|$ = $9$.
\end{itemize} 
\label{HF sphere example}
\end{example}

For any given non-negative integer $r$ and any given $b\in\mathcal{A}_q$, consider HF sequences $\a$ = $a_1a_2\ldots a_n$ and $\b$ = $b_1b_2\ldots b_n$ each of length $n$ such that $\b\in H_r(\a)$ and $b_n=b$.
Then, in the HF sphere $H_r(\a)$, the number of those HF sequences is $S_\a(r;b)$, i.e., \[S_\a(r;b) \overset{\Delta}{=} |\{\b:b_n=b \mbox{ for }\b\in H_r(\a)\}|,\] where $|S|$ is the size of the set $S$.
For $q$ = $4$, if $\mathcal{A}_4$ = $\{\alpha_1,\alpha_2,\alpha_3,\alpha_4\}$ then $\mathcal{C}_{4,2}$ = $\left\{\alpha_1\alpha_2,\alpha_1\alpha_3,\alpha_1\alpha_4,\alpha_2\alpha_1,\alpha_2\alpha_3\right.$, $\left.\alpha_2\alpha_4,\alpha_3\alpha_1,\alpha_3\alpha_2,\alpha_3\alpha_4,\alpha_4\alpha_1,\alpha_4\alpha_2,\alpha_4\alpha_3\right\}$.
Now, consider $H_2(\alpha_1\alpha_2)$ = $\left\{\alpha_2\alpha_1,\alpha_2\alpha_3,\alpha_2\alpha_4,\alpha_3\alpha_1,\alpha_3\alpha_4,\alpha_4\alpha_1,\alpha_4\alpha_3\right\}$.
Then, $S_{\alpha_1\alpha_2}(2,\alpha_1)$ = $3$, $S_{\alpha_1\alpha_2}(2,$ $\alpha_2)$ = $0$, $S_{\alpha_1\alpha_2}(2,\alpha_3)$ = $2$ and $S_{\alpha_1\alpha_2}(2,\alpha_4)$ = $2$.
Now, Proposition \ref{pro 1} is as follows.
\begin{proposition}
For any given integers $n$ ($\geq3$) and $r$, and a sequence $\a$ = $a_1a_2\ldots a_n\in\mathcal{C}_{q,n}$, the HF sphere size is 
\begin{equation}
    |H_r(\a)| = \sum_{b\in\mathcal{A}_q} S_\a(r;b).
    \label{HF sphere size value equation}
\end{equation}
\label{pro 1}
\end{proposition}

For any $(n,M,d)_q$ HF code $\mathcal{C}$ and codeword $\c\in\mathcal{C}$, the average size sum of HF spheres is 
\begin{equation}
\bar{U}_{HF}(\mathcal{C},r)\overset{\Delta}{=}\frac{1}{M}\sum_{\c\in\mathcal{C}}\sum_{i=0}^r|H_{i}(\c)|.
\label{average HF sphere size}
\end{equation}
For any $(n,M,d)_q$ HF code $\mathcal{C}$, the maximum size is \[A_q^{HF}(n,d)\overset{\Delta}{=}\max\limits_{\mathcal{C}}M.\]  
For any  $(n,M,d)_q$ HF code $\mathcal{C}$ with given $\bar{U}_{HF}(\mathcal{C},r)$, the maximum size is  
\begin{equation*}
    \begin{split}
        A_q^{HF}(n,d;&\bar{U}_{HF}(\mathcal{C},r)) \\ \overset{\Delta}{=} & \max\{M:\bar{U}_{HF}(\mathcal{C},r)\mbox{ are the same for all }\mathcal{C}\}.
    \end{split}
\end{equation*}
If sequences $\c_{max}$ and $\c_{min}$ are the given codewords of an $(n,M,d)_q$ HF code $\mathcal{C}$ $s.t.$ \\ 
$\mbox{\hspace{0.5cm}}W_{HF}(\c_{max},r)\overset{\Delta}{=}\max\limits_{\c\in\mathcal{C}}\sum_{i=0}^r|H_{i}(\c)|$, and \\ 
$\mbox{\hspace{0.5cm}}W_{HF}(\c_{min},r)\overset{\Delta}{=}\min\limits_{\c\in\mathcal{C}}\sum_{i=0}^r|H_{i}(\c)|$ then \\
$\mbox{\hspace{0.5cm}}A_q^{HF}(n,d,\c_{min})\overset{\Delta}{=}\max\{M:\c_{min}\in\mathcal{C}\}$, and \\
$\mbox{\hspace{0.5cm}}A_q^{HF}(n,d,\c_{max})\overset{\Delta}{=}\max\{M:\c_{max}\in\mathcal{C}\}$. \\
Now, consider sequences $\a_{max}$ and $\a_{min}$ in $\mathcal{C}_{q,n}$ such that 
\begin{equation*}
    \begin{split}
        S_{HF}(\a_{max},r) \overset{\Delta}{=}&\sum_{i=0}^r|H_{i}(\a_{max})| \\ = & \max\left\{\sum_{i=0}^r|H_{i}(\a)|:\a\in\mathcal{C}_{q,n}\right\},
    \end{split}
\end{equation*} 
and 
\begin{equation*}
    \begin{split}
        S_{HF}(\a_{min},r)\overset{\Delta}{=}&\sum_{i=0}^r|H_{i}(\a_{min})| \\ =&\min\left\{\sum_{i=0}^r|H_{i}(\a)|:\a\in\mathcal{C}_{q,n}\right\}.
    \end{split}
\end{equation*}
Also, note that $\c_{min}$ and $\c_{max}$ are the code property, and therefore, $S_{HF}(\a_{max},r)$ and $S_{HF}(\a_{min},r)$ depend on the code $\mathcal{C}$.

\section{Bounds on Homopolymer Free Codes}\label{Sec: HF Codes}
In this section, we have covered the properties of HF codes and established bounds on their size.
\subsection{Enumerating HF Sequences}
We have enumerated HF sequences for various lengths and alphabet sizes in this section.

Now, Lemma \ref{size C lemma} and Remark \ref{code rate remark} are as follows.
\begin{lemma}
For any given integer $n$, the number of distinct HF sequences each of length $n$ over the alphabet $\mathcal{A}_q$ is $q(q-1)^{n-1}$, i.e., the size of the set $\mathcal{C}_{q,n}$ is $|\mathcal{C}_{q,n}|=q(q-1)^{n-1}$.
\label{size C lemma}
\end{lemma}
\begin{IEEEproof}
For any HF sequence $\a$ = $a_1a_2\ldots a_n$ in $\mathcal{C}_{q,n}$, observe that $a_1\in\mathcal{A}_q$ and, for $i=2,3,\ldots,n$, $a_i\in\mathcal{A}_q\backslash\{a_{i-1}\}$. 
Thus, the size of $\mathcal{A}_q$ and $\mathcal{A}_q\backslash\{a_{i-1}\}$ are $q$ and $q-1$, respectively. 
Hence, the size $|\mathcal{C}_{q,n}|=q(q-1)^{n-1}$.
\end{IEEEproof}
\begin{remark}
For any given alphabet $\mathcal{A}_q$, the asymptotic code rate is 
\[\lim\limits_{n\to\infty}\frac{1}{n}\log_q|\mathcal{C}_{q,n}| = \lim\limits_{n\to\infty}\frac{1}{n}\log_qq(q-1)^{n-1}\rightarrow\log_q(q-1).\]
Therefore, the asymptotic code rate of any $(n,M,d)_q$ HF code over $\mathcal{A}_q$ cannot be more than $\log_q(q-1)$.
In general, note that $\lim\limits_{n\to\infty; q\to\infty}\frac{\log_q|\mathcal{C}_{q,n}|}{n}\rightarrow 1$, since $\lim\limits_{q\to\infty}(\log_q(q-1))\rightarrow 1$ and $\lim\limits_{n\to\infty}(1/n)\rightarrow 0$.
\label{code rate remark}
\end{remark}

\subsection{HF Sphere Size}
In this section, results on the size of an HF sphere are discussed using recurrence relations.

From Example \ref{HF sphere example}, for two HF sequences of the same length, one can observe that the sizes of two HF spheres may not be the same.
Also, $|H_1(\alpha_1\alpha_2\alpha_3)|<|H_1(\alpha_1\alpha_2\alpha_1)|$ where the size of the HF Sphere depends on the centres $\alpha_1\alpha_2\alpha_3$ and $\alpha_1\alpha_2\alpha_1$.
Thus, the HF sphere size depends on the pattern of the centre sequence. 
Therefore, unlike the general case, we obtained the HF sphere size $|H_{r}(\a)|$ from $S_r(\a,b)$ for any $b\in\mathcal{A}_q$ (Equation (\ref{HF sphere size value equation})) and  $S_r(\a,b)$ is obtained from recurrence relations.
In Proposition \ref{Initial Conditions 1 1}, we have given the value $S_{a_1}(1;b)$ for $r=1$ and $n=1$.
Then, we have obtained the value $S_{\a}(1;b)$ for $r=1$ and any integer $n$ in Lemma \ref{S1 lemma} with the initial conditions given in Proposition \ref{Initial Conditions 1 1}.
Further, we have given the value $S_{\a}(r;b)$ for any integer $n$ and $r$ ($\leq n$) in Lemma \ref{Sr lemma} with the initial conditions given in Lemma \ref{S1 lemma}.
So, Proposition \ref{Initial Conditions 1 1}, Lemma \ref{S1 lemma}, and Lemma \ref{Sr lemma} are as follows.
\begin{proposition}
For any $a_1,b\in\mathcal{A}_q$, \\ 
\[S_{a_1}(1;b) = 
    \begin{cases}
    0   & \mbox{for }a_1 = b \\
    1   & \mbox{otherwise}.
    \end{cases}
\]
\label{Initial Conditions 1 1}
\end{proposition}
\begin{lemma}
For any given integer $n$, and any given $b\in\mathcal{A}_q$, consider a sequence $\a$ = $a_1a_2\ldots a_n\in\mathcal{C}_{q,n}$ and the sub-sequences $\a(t)$ = $a_1a_2\ldots a_t$ for $1\leq t\leq n$.
Then, for $n\geq2$, 
\[
S_{\a(n)}(1;b) = 
    \begin{cases}
    \sum\limits_{c\in\mathcal{A}_q\backslash\{b\}}S_{\a(n-1)}(1;c)   & \mbox{for }a_n = b \\
    0 & \mbox{for }a_{n-1} = b \\
    1   & \mbox{otherwise}
    \end{cases}
\]
with the initial conditions as given in Proposition \ref{Initial Conditions 1 1}.
\label{S1 lemma}
\end{lemma}
\begin{IEEEproof}
For any $a_1\in\mathcal{A}_q$, one can find that the HF sphere with center $a_1$ and radius $1$ is $H_1(a_1)$ = $\mathcal{A}_q\backslash\{a_1\}$.
And therefore, the result on $S_{a_1}(1;b)$ follows for any $b\in\mathcal{A}_q$. 
Now, for any $\a(n)$ = $a_1a_2\ldots a_n$ in $\mathcal{C}_{q,n}$, recall that $a_i\neq a_{i+1}$ ($i$ = $1,2,\ldots,n-1$).
In particular, $a_{n-1}\neq a_n$.
For any $\b(n)$ = $b_1b_2\ldots b_n$ in the HF sphere $H_1(\a(n))$ with center $\a(n)$ and radius $1$, the Hamming distance $d(\a(n),\b(n))$ = $1$.
Now, there are two cases as follows.
\begin{itemize}
    \item \textbf{Case 1}: If $a_n=b_n$ then, from $a_{n-1}\neq a_n$, $a_{n-1}\neq b_n$.
    Therefore, from $d(\a(n),\b(n))$ = $1$, the Hamming distance $d(\a(n-1),\b(n-1))$ = $1$.
    Thus, for $a_n=b_n$, $S_{\a(n)}(1;b_n) = \sum_{c\in\mathcal{A}_q\backslash\{b_n\}}S_{\a(n-1)}(1;c)$. 
    \item \textbf{Case 2}: If $a_n\neq b_n$ then again there are two Sub-Cases as follows.
    \begin{itemize}
        \item \textbf{Sub-Case 1}: If $a_{n-1}=b_n$ then, from $a_{n-1}\neq a_n$, $a_n\neq b_n$.
        Therefore, from $d(\a(n),\b(n))$ = $1$, the Hamming distance $d(\a(n-1),\b(n-1))$ = $0$, and thus $a_{n-1}$ = $b_{n-1}$. 
        This imply that $b_{n-1}$ = $b_n$, but, the sequence $\b$ is an HF sequence, and therefore, contradicts $b_{n-1}$ = $b_n$ for this case.
        Hence, $S_{\a(n)}(1;b_n)$ = $0$ for $a_{n-1}=b_n$. 
        \item  \textbf{Sub-Case 2}: If $a_{n-1}\neq b_n$ then, from $d(\a(n),\b(n))$ = $1$, the Hamming distance $d(\a(n-1),\b(n-1))$ = $0$.
    Thus, for $a_{n-1}\neq b_n$, the sequence $a_1a_2\ldots a_{n-1}b_n$ belongs to the HF sphere $H_1(\a(n))$.
    So, $S_{\a(n)}(1;b_n)$ = $1$ for $a_n\neq b_n$, $a_{n-1}\neq b_n$. 
    \end{itemize}
\end{itemize}
Hence the result follows from all the cases for $b_n$ = $b$.
\end{IEEEproof}
\begin{lemma}
For any given integers $r$ and $n$, and any given $b\in\mathcal{A}_q$, consider a sequence $a_1a_2\ldots a_n\in\mathcal{C}_{q,n}$.
Then, for $2\leq n$ and $2\leq r\leq n$, 
\[
S_{\a(n)}(r;b)=
    \begin{cases}
    \sum\limits_{c\in\mathcal{A}_q\backslash\{b\}}S_{\a(n-1)}(r;c)    & \mbox{for }a_n = b \\
    \sum\limits_{c\in\mathcal{A}_q\backslash\{b\}}S_{\a(n-1)}(r-1;c)  & \mbox{otherwise},
    \end{cases}
\]
where the initial conditions $S_{\a(k)}(1;a_k)$ for $k$ = $1,2,\ldots,$ $n-r+1$ can be obtained from Lemma \ref{S1 lemma}. 
\label{Sr lemma}
\end{lemma}
\begin{IEEEproof}
For any integers $n$ and $r$ ($\leq n$), consider sequences $\a(n)$ = $a_1a_2\ldots a_n$ and $\b(n)$ = $b_1b_2\ldots b_n$ in $\mathcal{C}_{q,n}$ such that $\b(n)\in H_r(\a(n))$, and therefore, $d(\a(n),\b(n))$ = $r$.
Now, there are two cases that follow.
\begin{enumerate}
    \item If $a_n=b_n$ then $d(\a(n-1),\b(n-1))$ = $r$, and therefore, \[S_{\a(n)}(r;b_n) = \sum_{c\in\mathcal{A}_q\backslash\{b\}}S_{\a(n-1)}(r;c).\]
    \item If $a_n\neq b_n$ then $d(\a(n-1),\b(n-1))$ = $r-1$, and therefore, \[S_{\a(n)}(r;b_n) = \sum_{c\in\mathcal{A}_q\backslash\{b\}}S_{\a(n-1)}(r-1;c).\]
\end{enumerate}
Hence the result follows from both the cases for $b_n$ = $b$ with the initial conditions $S_{\a(k)}(1;a_k)$ for $k$ = $1,2,\ldots,n-r+1$ that can be obtained from Lemma \ref{S1 lemma}.
\end{IEEEproof}

Now, from Equation (\ref{HF sphere size value equation}), one can obtain the size of any HF Sphere for given a centre $\a$ of length $n$ and a radius $r$ using Lemma \ref{Sr lemma} in general.
For the particular case $r=1,2$, we have given the size of an HF Sphere in Lemma \ref{V 1 gen} and Lemma \ref{V 2 gen}, and for that we give Definition \ref{Chr Sq def} as follows.

\begin{definition}
For an integer $\ell$ $(1<\ell<n)$ and a sequence $\a$ = $a_1a_2\ldots a_n$ of length $n$ over $\mathcal{A}_q$, the binary sequence $\bar{\tau}^{(\ell)}(\a)$ = $\tau^{(\ell)}_1\tau^{(\ell)}_2\ldots\tau^{(\ell)}_n$ is called characteristic sequence of type $\ell$, where 
\[
\tau^{(\ell)}_{i+1} =
\begin{cases}
    1 & \mbox{for } a_i=a_{i+\ell} \mbox{ and }1\leq i\leq n-\ell \\
    0 & \mbox{otherwise.}
\end{cases}
\]
\label{Chr Sq def}
\end{definition}
For example, if $\a$ = $1213234$ in $\mathcal{C}_{4,7}$ then $\bar{\tau}^{(2)}(\a)$ = $0100100$ and $\bar{\tau}^{(3)}(\a)$ = $0010000$.
% Now, for the rest of the paper, we have considered $\tau^{(\ell)}_{p;q}$ = $\sum_{i=p}^q\tau^{(\ell)}_i$. 
Now, we have enumerated the HF sphere sizes $|H_{1}(\a)|$ and $|H_{2}(\a)|$ for any $\a\in\mathcal{C}_{q,n}$ in Lemma \ref{V 1 gen} and Lemma \ref{V 2 gen} as follows.
\begin{lemma}
For an HF sequence $\a$ = $a_1a_2\ldots a_n$ in $\mathcal{C}_{q,n}$, 
\begin{equation}
|H_{1}(\a)| = 2+n(q-3)+\sum_{i=2}^{n-1}\tau^{(2)}_i.
\label{eq: V 1 gen}
\end{equation}
\label{V 1 gen}
\end{lemma}
\begin{IEEEproof}
For any HF sequence $\a$ = $a_1a_2\ldots a_n$ in $\mathcal{C}_{q,n}$, consider $\b$ = $b_1b_2\ldots b_n$ in $\mathcal{C}_{q,n}$ such that $d(\a,\b)=1$. 
This implies that there is only one position $j$ such that $a_j\neq b_j$ and $a_i=b_i$ for $i\neq j$ and $i=1,2,\ldots,n$. 
Now, there are three cases as follows. 
\begin{itemize}
    \item \textbf{Case 1} ($j=1$): In this case, $b_1\neq a_1$ (from distance property) and $b_1\neq b_2=a_2$ (from HF sequence property), and thus, there exist $(q-2)$ HF sequences such that $d(\a,\b)=1$ and $j=1$.
    \item \textbf{Case 2} ($j=n$): In this case, $b_n\neq a_n$ (from distance property) and $b_n\neq b_{n-1}=a_{n-1}$ (from HF sequence property), and similar to Case 1, there exists $(q-2)$ HF sequences such that $d(\a,\b)=1$ and $j=n$.
    \item \textbf{Case 3} ($1<j<n$): In this case, $b_j\neq a_j$ (from distance property), $b_j\neq b_{j-1}=a_{j-1}$ and $b_j\neq b_{j+1}=a_{j+1}$ (from HF sequence property), and therefore, there exists $(q-3)$ HF sequences for $a_{j-1}\neq a_{j+1}$ and $(q-2)$ HF sequences for $a_{j-1}=a_{j+1}$
    Thus, for $1<j<n$, the number of HF sequences $\b$ such that $d(\a,\b)=1$ is \[\sum_{j=2}^{n-1}(q-3+\tau_j^{(2)})=(n-2)(q-3)+\sum_{j=2}^{n-1}\tau_j^{(2)}.\]
\end{itemize}
The result follows from adding the number of HF sequences enumerated in Case 1, Case 2 and Case 3.
\end{IEEEproof}
\begin{lemma}
For an HF sequence $\a$ = $a_1a_2\ldots a_n$ in $\mathcal{C}_{q,n}$, the HF sphere size $|H_{2}(\a)|$ is given in Equation (\ref{eq: V 2 gen}).
\begin{figure*}
    \begin{equation}
        \begin{split} 
    |H_{2}(\a)| = &  3(q-2)^2+2+(n-3)\left[(q-2)(q-3)+2\right] -2\sum_{k=2}^{n-1}\tau_k^{(2)}-\sum_{k=2}^{n-2}\tau_k^{(3)} \\     
        & +2(q-2)\sum_{j=3}^{n-1}\left((q-3)+\tau_j^{(2)}\right) +\sum_{i=2}^{n-3}\left(\left((q-3)-\tau_i^{(2)}\right)\cdot\sum_{j=i+2}^{n-1}\left((q-3)+\tau_j^{(2)}\right)\right) \\ 
        \end{split}
        \label{eq: V 2 gen}
    \end{equation}    
\end{figure*}
\label{V 2 gen}
\end{lemma}
\begin{IEEEproof}
For any sequence $\a = a_1a_2\ldots a_n$ in $\mathcal{C}_{q,n}$, consider $b=b_1b_2\ldots b_n$ such that $d(\a,\b)=2$. 
This implies that there are two positions $i$ and $j$ ($i<j$), where $a_i\neq b_i$ and $a_j\neq b_j$.
From combinatorics, if symbols $b_i$ and $b_j$ are at consecutive positions then there are $2(q-2)^2+2+(n-3)\left[(q-2)(q-3)+2\right]-2\sum_{k=2}^{n-1}\tau_k^{(2)}-\sum_{k=2}^{n-2}\tau_k^{(3)}$ sequences, and if symbols $b_i$ and $b_j$ are not at consecutive positions then there are $(q-2)^2+2\left[(q-2)\sum_{j=3}^{n-1}\left((q-3)+\tau_j^{(2)}\right)\right] +\sum_{i=2}^{n-3}\left(\left((q-3)-\tau_i^{(2)}\right)\cdot\sum_{j=i+2}^{n-1}\left((q-3)+\tau_j^{(2)}\right)\right)$ sequences such that $d(\a,\b)=2$.
Hence, the result can be obtained by adding these two cases.
For more details, please refer to Appendix \ref{App. V 2 gen proof}.
\end{IEEEproof}

From Equation (\ref{eq: V 1 gen}) and Equation (\ref{eq: V 2 gen}), for a given HF sequence $\a$ in $\mathcal{C}_{q,n}$, the HF sphere size depends on the number of ones in the $\bar{\tau}^{(2)}(\a)$ and $\bar{\tau}^{(3)}(\a)$.
In Proposition \ref{i+3 size} and Proposition \ref{i+2 size}, we have obtained HF sphere sizes with radius one and two for HF sequences $\a_{min}$ and $\a_{max}$ with specific patterns that minimize or maximize the HF sphere size as follows.
\begin{proposition}
    For $q\geq4$, consider $\a$ = $a_1a_2\ldots a_n$ ($=\a_{min}$) in $\mathcal{C}_{q,n}$ $s.t.$ $a_i = a_{i+3}$ for $i=1,2,\ldots,n-3$.
    Then,  \[|H_{1}(\a)| = (q-3)n+2,\] and  
        \[|H_{2}(\a)| = 
            \begin{cases}
                (q-1)(q-2)+1 & \mbox{for }n=2 \\
                \frac{1}{2}(n-4)(n-3)(q-3)^2+n(3q^2 & \\
                \hspace{0.7cm}-15q+19)-6q^2+33q-43 & \mbox{for }n\geq3.
            \end{cases}
        \] 
    \label{i+3 size}
\end{proposition}
\begin{proposition}
    For $q\geq3$, consider $\a$ = $a_1a_2\ldots a_n$ ($=\a_{max}$) in $\mathcal{C}_{q,n}$ $s.t.$ $a_i = a_{i+2}$ for $i=1,2,\ldots,n-2$.
    Then,  
    \begin{itemize}
        \item the HF Sphere size with $r=1$ is 
        \[
        |H_{1}(\a)| =
            \begin{cases}
                q-1 & \mbox{ for }n=1 \\
                n(q-2) & \mbox{ for }n\geq 2, \mbox{ and}
            \end{cases}
        \]
        \item the HF Sphere size with $r=2$ is 
        \[ |H_{2}(\a)| = 
            \begin{cases}
                q^2-3q+3 & \mbox{for }n=2 \\
                \frac{1}{2}(n-4)(n-3)(q-2)^2 & \\
                \hspace{0.8cm}+n(3q^2-13q+14) &  \\
                \hspace{1.3cm}-6q^2+27q-30 & \mbox{for }n\geq3.
            \end{cases}
        \]
    \end{itemize}
    \label{i+2 size}
\end{proposition}

\subsection{SP Bounds for HF Codes}
In this section, the SP bounds are calculated for HF codes in Theorem \ref{SP Bound Theorem 1}, Theorem \ref{SP Bound Theorem 2} and Theorem \ref{SP Bound Theorem 3}.
Before the SP bounds, we need to prove some results as given in Lemma \ref{bound 1} as follows.
\begin{lemma}
For any $(n,M,d)_q$ HF code $\mathcal{C}$ over $\mathcal{A}_q$, \[\sum_{\c\in\mathcal{C}}\sum_{r=0}^{\left\lfloor(d-1)/2\right\rfloor}|H_{r}(\c)|\leq q(q-1)^{n-1}.\]
\label{bound 1}
\end{lemma}
\begin{IEEEproof}
For any $(n,M,d)_q$ HF code $\mathcal{C}$, if $\c_1,\c_2\in\mathcal{C}$ such that $\c_1\neq\c_2$ then $H_{r_1}(\c_1)\cap H_{r_2}(\c_2)$ = $\emptyset$, where $r_1$ and $r_2$ are integers such that $0\leq r_1+r_2\leq d-1$.
\end{IEEEproof}
For given $n$, $q$ and $d$, if the maximum size is $A_q^{HF}(n,d)$ then the HF code with the size $A_q^{HF}(n,d)$ is denoted by $\mathcal{C}_{max}$. 
Now, Lemma \ref{bound 2} and Lemma \ref{bound inequality} are as follows.
\begin{lemma}
For any $(n,A_q^{HF}(n,d),d)_q$ HF code $\mathcal{C}_{max}$, \[\sum_{\c\in\mathcal{C}_{max}}\sum_{r=0}^{d-1}|H_{r}(\c)|\geq q(q-1)^{n-1}.\]
\label{bound 2}
\end{lemma}
\begin{IEEEproof}
Consider an $(n,A_q^{HF}(n,d),d)_q$ HF code $\mathcal{C}_{max}$ over $\mathcal{A}_q$. 
Then, for each $\a\in\mathcal{C}_{q,n}$, there exists at least one $\c\in\mathcal{C}_{max}$ such that $\c\in\bigcup_{r=0}^{d-1}H_r$.
Thus, the result follows from the set inclusion-exclusion principle.
\end{IEEEproof}
\begin{lemma}
For any integers $n$, $R$ and $r$ $(0\leq r\leq R\leq n)$, an $(n,M,d)_q$ HF code $\mathcal{C}$ over $\mathcal{A}_q$ follows inequality (\ref{HF inequality}).
\begin{figure*}
\begin{equation}
\sum_{r=0}^R|H_{r}(\a_{min})|\leq\sum_{r=0}^R|H_{r}(\c_{min})|\leq\frac{1}{M}\sum_{\c\in\mathcal{C}}\sum_{r=0}^R|H_{r}(\c)|\leq\sum_{r=0}^R|H_{r}(\c_{max})|\leq\sum_{r=0}^R|H_{r}(\a_{max})|
\label{HF inequality}
\end{equation}
\end{figure*}
\label{bound inequality}
\end{lemma}
\begin{IEEEproof}
From Equation (\ref{average HF sphere size}), \[\frac{1}{M}\sum_{\c\in\mathcal{C}}\sum_{r=0}^R|H_{r}(\c)| = \bar{U}_{HF}(\mathcal{C},R).\] 
And therefore, from the definitions of $\c_{max}$ and $\c_{min}$, \[\sum_{r=0}^R|H_{r}(\c_{min})|\leq\bar{U}_{HF}(\mathcal{C},R)\leq\sum_{r=0}^R|H_{r}(\c_{max})|.\]
Again from the definitions of $\a_{max}$ and $\a_{min}$, \[\sum_{r=0}^R|H_{r}(\a_{min})|\leq\sum_{r=0}^R|H_{r}(\c_{min})|\] and \[\sum_{r=0}^R|H_{r}(\c_{max})|\leq\sum_{r=0}^R|H_{r}(\a_{max})|.\]
\end{IEEEproof}

For any HF codes over $\mathcal{A}_4$, upper bounds are given in Theorem \ref{SP Bound Theorem 1}, Theorem \ref{SP Bound Theorem 2} and Theorem \ref{SP Bound Theorem 3} as follows.
\begin{theorem}[Upper Bound 1]
For any HF code, \[ A_q^{HF}(n,d)\leq\left\lfloor\frac{q(q-1)^{n-1}}{S_{HF}(\a_{min},\left\lfloor\frac{d-1}{2}\right\rfloor)}\right\rfloor.\]
\label{SP Bound Theorem 1}
\end{theorem}
\begin{IEEEproof} 
From Lemma \ref{bound inequality}, for $R$ = $\left\lfloor\frac{d-1}{2}\right\rfloor$, one can obtain \[\sum_{r=0}^{\left\lfloor\frac{d-1}{2}\right\rfloor}|H_{r}(\a_{min})|\leq\frac{1}{M}\sum_{\c\in\mathcal{C}}\sum_{r=0}^{\left\lfloor\frac{d-1}{2}\right\rfloor}|H_{r}(\c)|,\] and therefore, from Lemma \ref{bound 1}, the result follows for the $(n,A_q^{HF}(n,d),d)_q$ HF code.
\end{IEEEproof}

\begin{theorem}[Upper Bound 2]
For any HF code $\mathcal{C}$ such that $\c_{min}\in\mathcal{C}$, 
\[A_q^{HF}(n,d,\c_{min})\leq\left\lfloor\frac{q(q-1)^{n-1}}{W_{HF}(\c_{min},\left\lfloor\frac{d-1}{2}\right\rfloor)}\right\rfloor.\]
\label{SP Bound Theorem 2}
\end{theorem}
\begin{IEEEproof} 
From Lemma \ref{bound inequality}, for $R$ = $\left\lfloor\frac{d-1}{2}\right\rfloor$, one can obtain \[\sum_{r=0}^{\left\lfloor\frac{d-1}{2}\right\rfloor}|H_{r}(\c_{min})|\leq\frac{1}{M}\sum_{\c\in\mathcal{C}}\sum_{r=0}^{\left\lfloor\frac{d-1}{2}\right\rfloor}|H_{r}(\c)|,\] and therefore, from Lemma \ref{bound 1}, the result follows for the $(n,A_q^{HF}(n,d,\c_{min}),d)_q$ HF code.
\end{IEEEproof}

\begin{theorem}[Upper Bound 3]
For any HF code $\mathcal{C}$ with given $\bar{U}_{HF}(\mathcal{C},\left\lfloor\frac{d-1}{2}\right\rfloor)$, \[A_q^{HF}(n,d;\bar{U}_{HF}(\mathcal{C},\lfloor(d-1)/2\rfloor))\leq\left\lfloor\frac{q(q-1)^{n-1}}{\bar{U}_{HF}(\mathcal{C},\left\lfloor\frac{d-1}{2}\right\rfloor)}\right\rfloor\]
\label{SP Bound Theorem 3}
\end{theorem}
\begin{IEEEproof}
For given $r$, observe $\frac{1}{M}\sum_{\c\in\mathcal{C}}\sum_{i=0}^r|H_{i}(\c)|$ = $\bar{U}_{HF}(\mathcal{C},r)$, and thus, from Lemma \ref{bound 1}, the result follows for the $(n,A_q^{HF}(n,d;\bar{U}_{HF}(\mathcal{C},\lfloor(d-1)/2\rfloor)),d)_q$ HF code.
\end{IEEEproof}

Note that Upper Bound 1 is the upper bound for any $(n, M, d)_q$ HF code, Upper Bound 2 is the upper bound among those $(n,M,d)_q$ HF codes that has the same $\c_{min}$, and Upper Bound 3 is the upper bound among those $(n, M, d)_q$ HF codes that has the same $\bar{U}_{HF}(\mathcal{C},r)$.

\subsection{GV Bounds  for HF Codes}
In this section, lower bounds are calculated for HF codes in Theorem \ref{GV Bound Theorem 1}, Theorem \ref{GV Bound Theorem 2} and Theorem \ref{GV Bound Theorem 3} as follows.
\begin{theorem}[Lower Bound 1]
For any HF code, \[A_q^{HF}(n,d)\geq\left\lceil\frac{q(q-1)^{n-1}}{S_{HF}(\a_{max},d-1)}\right\rceil.\]
\label{GV Bound Theorem 1}
\end{theorem}
\begin{IEEEproof}
From Lemma \ref{bound inequality}, for $R$ = $d-1$, one can obtain $\sum_{r=0}^{d-1}|H_{r}(\a_{max})|\geq\frac{1}{M}\sum_{\c\in\mathcal{C}}\sum_{r=0}^{d-1}|H_{r}(\c)|$, and therefore, from Lemma \ref{bound 2}, the result follows.
\end{IEEEproof}

\begin{theorem}[Lower Bound 2]
For any HF code $\mathcal{C}$ such that $\c_{max}\in\mathcal{C}$, \\ 
\[A_q^{HF}(n,d,\c_{max})\geq\left\lfloor\frac{q(q-1)^{n-1}}{W_{HF}(\c_{max},d-1)}\right\rfloor.\]
\label{GV Bound Theorem 2}
\end{theorem}
\begin{IEEEproof}
From Lemma \ref{bound inequality}, for $R$ = $d-1$, one can obtain $\sum_{r=0}^{d-1}|H_{r}(\c_{max})|\geq\frac{1}{M}\sum_{\c\in\mathcal{C}}\sum_{r=0}^{d-1}|H_{r}(\c)|$, and therefore, from Lemma \ref{bound 2}, the result follows for the $(n,A_q^{HF}(n,d,\c_{max}),d)_q$ HF code.
\end{IEEEproof}

\begin{theorem}[Lower Bound 3]
For any HF code $\mathcal{C}$  with given $\bar{U}_{HF}(\mathcal{C},d-1)$, \[A_q^{HF}(n,d;\bar{U}_{HF}(\mathcal{C},d-1))\geq\left\lfloor\frac{q(q-1)^{n-1}}{\bar{U}_{HF}(\mathcal{C},d-1)}\right\rfloor.\]
\label{GV Bound Theorem 3}
\end{theorem}
\begin{IEEEproof}
For given $r$, observe $\frac{1}{M}\sum_{\c\in\mathcal{C}}\sum_{i=0}^r|H_{i}(\c)|$ = $\bar{U}_{HF}(\mathcal{C},r)$, and thus, from Lemma \ref{bound 2}, the result follows for the $(n,A_q^{HF}(n,d;\bar{U}_{HF}(\mathcal{C},d-1)),d)_q$ HF code.
\end{IEEEproof}

Note that Lower Bound 1 is the lower bound for any $(n, M, d)_q$ HF code, Lower Bound 2 is the lower bound among those $(n, M, d)_q$ HF codes that has the same $\c_{max}$, and Lower Bound 3 is the lower bound among those $(n, M, d)_q$ HF codes that has the same $\bar{U}_{HF}(\mathcal{C},r)$.

\begin{figure}
\includegraphics[scale=0.65]{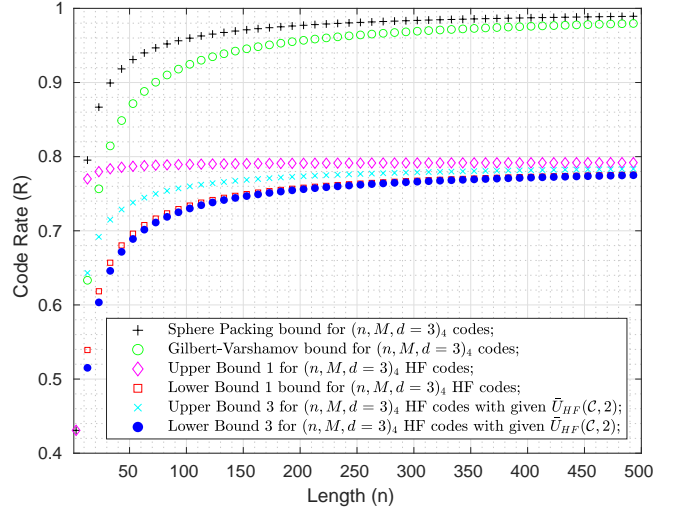}
\caption{Upper (SP) and lower (GV) bounds for codes over an alphabet $\mathcal{A}_4$ are plotted for $n=3,4,\ldots,500$, and $d=3$.}
\label{Plots}
\end{figure}

\section{Bounds on Homopolymer Free DNA Codes}\label{Sec: HF DNA Codes}
For $q$ = $4$, consider the alphabet $\mathcal{A}_4$ = $\{\alpha_1,\alpha_2,\alpha_3,\alpha_4\}$.
Then, the alphabet $\mathcal{A}_4$ corresponds to DNA alphabet $\{A,C,G,T\}$.
From Remark \ref{code rate remark}, the code rate of any HF DNA code cannot be more than $\log_43\approx0.79248$.

% For $q=4$, Proposition \ref{HF sphere Bound DNA}, Proposition \ref{UB Ave DNA}, Proposition \ref{HF GV Bound DNA} and Proposition \ref{LB Ave DNA} are obtained from  Theorem \ref{SP Bound Theorem 1}, Theorem \ref{SP Bound Theorem 3}, Theorem \ref{GV Bound Theorem 1} and Theorem \ref{GV Bound Theorem 3} as follows.

For $q=4$, Proposition \ref{HF sphere Bound DNA}, Proposition \ref{HF sphere Bound DNA 1}, Proposition \ref{UB Ave DNA}, Proposition \ref{HF GV Bound DNA}, Proposition \ref{HF GV Bound DNA 1} and Proposition \ref{LB Ave DNA} are obtained from Theorem \ref{SP Bound Theorem 1}, Theorem \ref{SP Bound Theorem 2}, Theorem \ref{SP Bound Theorem 3}, Theorem \ref{GV Bound Theorem 1}, Theorem \ref{GV Bound Theorem 2} and Theorem \ref{GV Bound Theorem 3} as follows.
\begin{proposition}[Upper Bound 1]
For any HF DNA code, \[A_4^{HF}(n,d)\leq\left\lfloor\frac{4\cdot3^{n-1}}{S_{HF}(\a_{min},\left\lfloor\frac{d-1}{2}\right\rfloor)}\right\rfloor.\]
\label{HF sphere Bound DNA}
\end{proposition}
\begin{proposition}[Upper Bound 2] For any HF DNA code $\mathcal{C}$ with given $\c_{min}\in\mathcal{C}$,  \[A_4^{HF}(n,d,\c_{min})\leq\left\lfloor\frac{4\cdot3^{n-1}}{W_{HF}(\c_{min},\left\lfloor\frac{d-1}{2}\right\rfloor)}\right\rfloor.\]
\label{HF sphere Bound DNA 1}
\end{proposition}
\begin{proposition}[Upper Bound 3] For any HF code $\mathcal{C}$ with given $\bar{U}_{HF}(\mathcal{C},\left\lfloor\frac{d-1}{2}\right\rfloor)$, 
\[A_4^{HF}(n,d;\bar{U}_{HF}(\mathcal{C},r))\leq\left\lfloor\frac{4\cdot3^{n-1}}{\bar{U}_{HF}(\mathcal{C},\left\lfloor\frac{d-1}{2}\right\rfloor)}\right\rfloor.\]
\label{UB Ave DNA}
\end{proposition}

\begin{proposition}[Lower Bound 1]
For any HF DNA code, \[A_4^{HF}(n,d)\geq\left\lceil\frac{4\cdot3^{n-1}}{S_{HF}(\a_{max},d-1)}\right\rceil.\]
\label{HF GV Bound DNA}
\end{proposition}
\begin{proposition}[Lower Bound 2] For any HF DNA code $\mathcal{C}$ with given $\c_{max}\in\mathcal{C}$,
\[A_4^{HF}(n,d,\c_{max})\geq\left\lceil\frac{4\cdot3^{n-1}}{W_{HF}(\c_{max},d-1)}\right\rceil.\]
\label{HF GV Bound DNA 1}
\end{proposition}
\begin{proposition}[Lower Bound 3] For any HF code $\mathcal{C}$  with given $\bar{U}_{HF}(\mathcal{C},d-1)$,
\[A_4^{HF}(n,d;\bar{U}_{HF}(\mathcal{C},d-1))\geq\left\lceil\frac{4\cdot3^{n-1}}{\bar{U}_{HF}(\mathcal{C},d-1)}\right\rceil.\]
\label{LB Ave DNA}
\end{proposition}

% \begin{proposition}[Upper Bound 1]
% For any HF DNA code, $A_4^{HF}(n,d)\leq\lfloor4\cdot3^{n-1}/S_{HF}(\a_{min},\left\lfloor\frac{d-1}{2}\right\rfloor)$.
% \label{HF sphere Bound DNA}
% \end{proposition}
% % \begin{IEEEproof}
% % The proof follows from Theorem \ref{SP Bound Theorem 1} for alphabet size four.
% % \end{IEEEproof}
% \begin{proposition}[Upper Bound 2] \\
%  $A_4^{HF}(n,d,\c_{min})\leq\lfloor4\cdot3^{n-1}/W_{HF}(\c_{min},\left\lfloor\frac{d-1}{2}\right\rfloor)$.
% \label{HF sphere Bound DNA 1}
% \end{proposition}
% % \begin{IEEEproof}
% % The proof follows from Theorem \ref{SP Bound Theorem 2} for alphabet size four. 
% % \end{IEEEproof}
% \begin{proposition}[Upper Bound 3] For $r=\lfloor(d-1)/2\rfloor)$, 
% $A_m^{HF}(n,d;\bar{U}_{HF}(\mathcal{C},r)\leq\lfloor4\cdot3^{n-1}/\bar{U}_{HF}(\mathcal{C},r)\rfloor$.
% \label{UB Ave DNA}
% \end{proposition}
% % \begin{IEEEproof}
% % The proof follows from Theorem \ref{SP Bound Theorem 3} for alphabet size four.
% % \end{IEEEproof}

% \begin{proposition}[Lower Bound 1]
% For any HF DNA code, $A_4^{HF}(n,d)\geq\lfloor4\cdot3^{n-1}/S_{HF}(\a_{max},d-1)\rfloor$.
% \label{HF GV Bound DNA}
% \end{proposition}
% % \begin{IEEEproof}
% % The proof follows from Theorem \ref{GV Bound Theorem 1} for alphabet size four.
% % \end{IEEEproof}

\begin{table}
\caption{Bounds on the maximum size of HF codes with $q=4$}
\begin{center}
\begin{tabular}{l|c|c|c|c|c|c|c|c|c|}
 \cline{2-10}
 & $d$ &\multicolumn{8}{|c|}{$n$} \\  \cline{3-10}
 &      & 1     & 2    & 3    & 4   & 5   & 6   & 7    & 8 \\ \hline
\multicolumn{1}{|l|}{Upper}  & 1   & 4     & 12   & 36   & 108 & 324 & 972 & 2916 & 8748 \\ \cline{2-10}
 \multicolumn{1}{|l|}{Bound 1}  & 2   & -     & 12   & 36   & 108 & 324 & 972 & 2916 & 8748 \\ \cline{2-10}
 \multicolumn{1}{|l|}{(Propos-}  & 3   & -     & -    & 6    & 15  & 40  & 108 & 291  & 795 \\ \cline{2-10}
 \multicolumn{1}{|l|}{ition \ref{HF sphere Bound DNA})} & 4   & -     & -    & -    & 15  & 8   & 108 & 291  & 795 \\ \cline{2-10}
 \multicolumn{1}{|l|}{} & 5   & -     & -    & -    & -   & 8   & 20  & 50   & 124 \\ \hline
 \multicolumn{1}{|l|}{Upper} & 1   & 4     & 12   & 36   & 108 & 324 & 972 & 2916 & 8748 \\ \cline{2-10}
 \multicolumn{1}{|l|}{Bound 3} & 2   & -     & 12   & 36   & 108 & 324 & 972 & 2916 & 8748 \\ \cline{2-10}
 \multicolumn{1}{|l|}{(Propos-} & 3   & -     & -    & 5    & 14  & 36  & 94  & 249  & 672  \\ \cline{2-10}
 \multicolumn{1}{|l|}{ition \ref{UB Ave DNA})} & 4   & -     & -    & -    & 14  & 36  & 94  & 249  & 672  \\ \cline{2-10}
 \multicolumn{1}{|l|}{}& 5   & -     & -    & -    & -   & 7   & 17  & 41   & 101  \\ \hline\hline
 \multicolumn{1}{|l|}{Lower} & 1   & 4     & 12   & 36   & 108 & 324 & 972 & 2916 & 8748   \\ \cline{2-10}
 \multicolumn{1}{|l|}{Bound 1} & 2   & -     & 3    & 6    & 12  & 30  & 75  & 195  & 515    \\ \cline{2-10}
 \multicolumn{1}{|l|}{(Pro. \ref{HF GV Bound DNA})} & 3   & -     & -    & 2    & 4   & 7   & 15  & 33   & 74     \\ \hline
 \multicolumn{1}{|l|}{Lower} & 1   & 4     & 12   & 36   & 108 & 324 & 972 & 2916 & 8748   \\ \cline{2-10}
 \multicolumn{1}{|l|}{Bound 3} & 2   & -     & 3    & 6    & 15  & 36  & 95  & 250  & 673    \\ \cline{2-10}
 \multicolumn{1}{|l|}{(Pro. \ref{LB Ave DNA})} & 3   & -     & -    & 2    & 4   & 8   & 18  & 42   & 102    \\ \hline
\end{tabular}
\end{center}
\label{Table}
\end{table}

\section{Discussions and Comparisons}\label{Sec: Discussion}
In Table \ref{Table}, we obtained the upper bounds on the maximum size  $A_4^{HF}(n,d)$ of HF code for $q=4$ (i.e., HF DNA codes) with $n=1,2,\ldots,8$ and $d=1,2,\ldots,\min\{5,n\}$ in first ten rows.
In this table, values in the first five rows values are Upper Bound 1 and are obtained from Proposition \ref{HF sphere Bound DNA} with the HF sphere size as given in Proposition \ref{i+3 size} for $q=4$.
The values in the next five rows values are the Upper Bound 3, and obtained from Proposition \ref{UB Ave DNA} with the average HF sphere size sum $\bar{U}_{HF}(\mathcal{C},r)$ = $\frac{1}{4\cdot3^{n-1}}\sum_{\c\in\mathcal{C}_{4,n}}|H_{r}(\c)|$ for $r=1,2$.
Then, we listed the lower bounds on the maximum size  $A_4^{HF}(n,d)$ of HF code for $q=4$ (i.e., HF DNA codes) with $n=1,2,\ldots,8$ and $d=1,2,\ldots, \min\{3,n\}$ in the remaining six rows of the same table.
In this table, values in the first three rows of the remaining six rows are Lower Bound 1 and are obtained from Proposition \ref{HF GV Bound DNA} with the HF sphere size as given in Proposition \ref{i+2 size} for $q=4$.
The rest three-row values are the Lower Bound 3, tighter than Lower Bound 1, and obtained from Proposition \ref{LB Ave DNA}, where the average HF sphere size sum is the same as the average size sum considered in Table \ref{Table}.
%%%%%%%%%%%%%%%%%%%%%%%%%%%%%

Further, in Fig. \ref{Plots}, bounds on various HF codes are studied for $q=4$ (i.e., HF DNA codes) and $n=3,4,\ldots,500$. 
In this figure, the black `$+$' and green `\tikz\draw[green] (0,0) circle (.4ex);' curves represent SP and GV bounds on quaternary codes (i.e., DNA codes), and the curves are obtained from inequality (\ref{SP bound Gen}) and from inequality (\ref{GV bound Gen}) with $V_q(n)=q^n$ and $V_q(n,r) = \binom{n}{r}(q-1)^r$, respectively.
Further, magenta `\textcolor{magenta}{$\diamondsuit$}' and red `\textcolor{red}{$\square$}' curves represent Upper Bound 1 and Lower Bound 1 on HF codes for $q=4$ (i.e., HF DNA codes), and the curves are obtained from Proposition \ref{HF sphere Bound DNA} and Proposition \ref{HF GV Bound DNA}, respectively.
Again, the cyan `\textcolor{cyan}{$\times$}' and blue `\tikz\draw[blue,fill=blue] (0,0) circle (.4ex);' curves in the figure represent Upper Bound 3 and Lower Bound 3 on HF codes for $q=4$ (i.e., HF DNA codes) with $\bar{U}_{HF}(\mathcal{C},2)$ = $12$ for $n=2$, $\bar{U}_{HF}(\mathcal{C},2)$ = $59/3$ for $n=3$, $\bar{U}_{HF}(\mathcal{C},2)$ = $\frac{8}{9}(n+2)^2+\frac{235}{9}$ for $n\geq4$, and also, the curves are obtained from Proposition \ref{UB Ave DNA} and Proposition \ref{LB Ave DNA}, respectively.
Note that all the curves obtained for various HF codes in Fig. \ref{Plots} are bounded above by $0.79248$ as given in Section \ref{Sec: HF DNA Codes}, and also, all the HF codes with $q=4$ (i.e., HF DNA codes) follow Upper Bound 1 and Lower Bound 1 in Fig. \ref{Plots} and Table \ref{Table}.

In this paper, Upper Bound 1 and Lower Bound 1 are the bounds on HF codes in general. 
Upper bound 2 and Lower Bound 2 are the bounds on the HF codes families that have the same $W_{HF}(\c_{max},r)$ and $W_{HF}(\c_{min},r)$, respectively.
Further, Upper Bound 3 and Lower Bound 3 are the bounds on the HF codes family that has the same $\bar{U}_{HF}(\mathcal{C},r)$. 
Note that the bounds discussed in this paper are for the codes where all the codewords are free from homopolymers of run-length $\ell\geq2$. 
If we limit the homopolymer property up to a given run-length $t$ (i.e. $\ell\leq t$) then the lower bounds obtained in the paper are also the lower bound for the codes where all the codewords are free from homopolymers of run-length $\ell\geq t$.

\bibliographystyle{IEEEtran} 
\bibliography{DNA}

\appendices

\section{Proof of Lemma \ref{V 2 gen}}
\label{App. V 2 gen proof}
For sequence $\a = a_1a_2\ldots a_n$ in $\mathcal{C}_{q,n}$, consider $b=b_1b_2\ldots b_n$ such that $d(\a,\b)=2$. 
This implies that there are two positions $i$ and $j$ ($i<j$), where $a_i\neq b_i$, $a_j\neq b_j$ and $a_k=b_k$ for $i\neq k$, $j\neq k$ and $k=1,2,\ldots,n$.
Now, there are two cases as discussed in the following two sections.
\subsection{Symbols $b_i$ and $b_j$ are at consecutive positions:} 
In this case, we have considered $j=i+1$. 
Now, there are again three cases as follows. 
\subsubsection{\textbf{Case 1} ($a_1\neq b_1$ and $a_2\neq b_2$)}
In this case, $i=1$ and $j=2$.
From HF sequence property, recall $a_1\neq a_2$, $a_2\neq a_3$, $b_1\neq b_2$ and $b_2\neq b_3$. 
The number of sequences $\b$ is the sum of the following two Sub-Cases.
\begin{enumerate}
    \item[i.] \textbf{Sub-Case 1} ($a_1=a_3$): 
    The Sub-Case follows from the following two parts. 
    \begin{itemize}
        \item \textbf{Part 1} ($b_1=a_2$): In this part, there is one option for $b_1$ and $(q-2)$ options for $b_2$.
        Therefore, for the particular Sub-Case, there are $1\cdot(q-2)$ options for the sequence $\b$.
        \item \textbf{Part 2} ($b_1\neq a_2$): In this part, then there are $(q-2)$ options for $b_1$ and again $(q-2)$ options for $b_2$.
        Therefore, for the particular Sub-Case, there are $(q-2)\cdot(q-3)$ options for the sequence $\b$.
    \end{itemize}
    Hence, for Sub-Case 1 ($i.e.\ a_1=a_3$), the number of sequences $\b$ is the sum of both the parts, $i.e.$, 
    $\#\b$ = $1\cdot(q-2)+(q-2)\cdot(q-3)$ = $(q-2)^2$. 
    \item[ii.] \textbf{Sub-Case 2} ($a_1\neq a_3$): 
    From HF sequence property, recall $a_1\neq a_2$ and $a_2\neq a_3$. 
    The Sub-Case follows from the following three parts. 
    \begin{itemize}
        \item \textbf{Part 1} ($b_1=a_2$): In this part, there is one option for $b_1$ and $(q-2)$ options for $b_2$.
        Therefore, for the particular Sub-Case, there are $1\cdot(q-2)$ options for the sequence $\b$.
        \item \textbf{Part 2} ($b_1=a_3$): In this part, then there is one option for $b_1$ and $(q-2)$ options for $b_2$.
        Therefore, for the particular Sub-Case, there are $1\cdot(q-2)$ options for the sequence $\b$.
        \item \textbf{Part 3} ($b_1\neq a_2$ and $b_1\neq a_3$): In this part, there are $(q-3)$ options for $b_1$ and again $(q-3)$ options for $b_2$.
        Therefore, for the particular Sub-Case, there are $(q-3)\cdot(q-3)$ options for the sequence $\b$.
    \end{itemize}
    Hence, for Sub-Case 2 ($i.e.\ a_1=a_3$), the number of sequences $\b$ is the sum of Part 1, Part 2 and Part 3, $i.e.$, 
    $\#\b$ = $1\cdot(q-2)+1\cdot(q-2)+(q-3)\cdot(q-3)$ \\ $\mbox{\hspace{0.65cm}}$= $(q-2)^2+1$.
\end{enumerate}
Thus, from Sub-Case 1 and Sub-Case 2, 
\[ \#\b =
\begin{cases}
    (q-2)^2+1 & \mbox{for } a_1=a_3 \\
    (q-2)^2 & \mbox{for } a_1\neq a_3, 
\end{cases}
\]
Therefore, from Definition \ref{Chr Sq def}, the number of sequences $\b$ such that $a_1\neq b_1$ and $a_2\neq b_2$ is 
\begin{equation}
    \#\b=(q-2)^2+1-\tau_2^{(2)}
    \label{V 2 eq1}
\end{equation}
\subsubsection{\textbf{Case 2} ($a_{n-1}\neq b_{n-1}$ and $a_n\neq b_n$)}
In this case, we have considered $i=n-1$ and $j=n$.
Similar to Case 1 as discussed above, the number of sequences $\b$ such that $a_{n-1}\neq b_{n-1}$ and $a_n\neq b_n$ is 
\begin{equation}
    \#\b=(q-2)^2+1-\tau_{n-1}^{(2)}
    \label{V 2 eq2}
\end{equation} 
\subsubsection{\textbf{Case 3} ($a_i\neq b_i$ and $a_{i+1}\neq b_{i+1}$ for $2\leq i\leq n-2$)}
    Similar to Case 1 ($a_1\neq b_1$ and $a_2\neq b_2$), we can enumerate the number of sequences $\b$ with $a_i\neq b_i$ and $a_{i+1}\neq b_{i+1}$ for $i=2,3,\ldots,n-2$  as following 
    \begin{equation}
        \#b=(q-2)(q-3)+2-\tau_2^{(2)}-\tau_3^{(2)}-\tau_2^{(3)}\mbox{ for }i=2 \label{eqn:line-1}
    \end{equation}
    \begin{equation}
        \#b=(q-2)(q-3)+2-\tau_3^{(2)}-\tau_4^{(2)}-\tau_3^{(3)}\mbox{ for }i=3 \label{eqn:line-2}
    \end{equation}
    % \begin{equation}
    %     \#b=(q-2)(q-3)+2-\tau_4^{(2)}-\tau_5^{(2)}-\tau_4^{(3)}\mbox{ for }i=4 \label{eqn:line-3}
    % \end{equation}
    \begin{equation*}
        \vdots
    \end{equation*}
    \begin{equation}
        \#b=(q-2)(q-3)+2-\tau_k^{(2)}-\tau_{k+1}^{(2)}-\tau_k^{(3)}\mbox{ for }i=k \label{eqn:line-4}
    \end{equation}
    \begin{equation*}
        \vdots
    \end{equation*}
    \begin{equation}
        \#b=(q-2)(q-3)+2-\tau_{n-2}^{(2)}-\tau_{n-1}^{(2)}-\tau_{n-2}^{(3)}\mbox{ for }i=n-2 \label{eqn:line-5}
    \end{equation}
    From adding Equation (\ref{eqn:line-1}) to Equation (\ref{eqn:line-5}), one can get the total number of sequences $\b$ with $a_i\neq b_i$ and $a_{i+1}\neq b_{i+1}$ for $i=2,3,\ldots,n-2$ is     
    \begin{equation*}
    \begin{split}
       \#b= &\sum_{k=2}^{n-2}\left((q-2)(q-3)+2-\tau_k^{(2)}-\tau_{k+1}^{(2)}-\tau_k^{(3)}\right) \\     
       = &(n-3)\left((q-2)(q-3)+2\right)-\tau_2^{(3)}-\tau_2^{(2)} \\ 
       &\hspace{3cm}-\tau_{n-1}^{(2)}-\sum_{k=3}^{n-2}\left(2\tau_k^{(2)}+\tau_k^{(3)}\right) \\        
    \end{split}
    \end{equation*}
    
    Hence, for given an HF sequence $\a$, the total number of HF sequences $\b$ such that $d(\a,\b)=2$ and both the HF sequences differ at consecutive locations is the sum of sequences enumerated in Case 1, Case 2 and Case 3, and the sum is  
    \begin{equation}
        \begin{split}       
       \#b= & 2(q-2)^2+2+(n-3)\left[(q-2)(q-3)+2\right] \\
        &\hspace{4cm}-2\sum_{k=2}^{n-1}\tau_k^{(2)}-\sum_{k=2}^{n-2}\tau_k^{(3)} \\  
        \end{split}
        \label{eq: V 2 1st part}
    \end{equation}

\subsection{$b_i$ and $b_j$ are not the symbols at consecutive positions:} 
In this case, we have considered $j=i+1$.
WLOG, let $j>i+1$, where $a_i\neq b_i$ and $a_j\neq b_j$. 
from HF sequence property, recall that $a_{i-1}\neq a_i$, $a_i\neq a_{i+1}$, $a_{j-1}\neq a_j$ and $a_j\neq a_{j+1}$, and also $b_{i-1}\neq b_i$, $b_i\neq b_{i+1}$, $b_{j-1}\neq b_j$ and $b_j\neq b_{j+1}$. 
From the Hamming distance property, $a_{i-1}\neq b_{i-1}$, $a_{i+1}\neq b_{i+1}$, $a_{j-1}\neq b_{j-1}$ and $a_{j+1}\neq b_{j+1}$.
There are four cases as follows.
\subsubsection{\textbf{Case 1} ($a_1\neq b_1$ and $a_n\neq b_n$)}
In this case, we have considered $i=1$ and $j=n$.
From Case 1 and Case 2 as discussed in the proof of Lemma \ref{V 1 gen}, there are $(q-2)$ options for both the symbols $b_1$ and $b_n$ for $n>2$.
Hence, there are $(q-2)^2$ HF sequences such that $a_1\neq b_1$ and $a_n\neq b_n$.
\subsubsection{\textbf{Case 2} ($a_1\neq b_1$ and $a_j\neq b_j$ for $j=3,4,\ldots,n-1$)}
In this case, we have considered $i=1$ and $j=3,4,\ldots,n-1$. 
From Case 1 and Case 3 as discussed in the proof of Lemma \ref{V 1 gen}, there are $(q-2)$ options for the symbols $b_1$ and, for given $j$, there are $(q-3)+\tau_j^{(2)}$ options for the symbols $b_j$.
Hence, there are $(q-2)\sum_{j=3}^{n-1}\left((q-3)+\tau_j^{(2)}\right)$ HF sequences such that $a_1\neq b_1$ and $a_j\neq b_j$ for $j=3,4,\ldots,n-1$.
\subsubsection{\textbf{Case 3} ($a_n\neq b_n$ and $a_i\neq b_i$ for $i=2,3,\ldots,n-2$)}
In this case, we have considered $i=2,3,\ldots,n-2$ and $j=n$.
Similar to Case 2, there are $(q-2)\sum_{i=3}^{n-1}\left((q-3)+\tau_i^{(2)}\right)$ HF sequences such that $a_n\neq b_n$ and $a_i\neq b_i$ for $i=2,3,\ldots,n-2$.
\subsubsection{\textbf{Case 4} ($a_i\neq b_i$ and $a_j\neq b_j$ for $i=2,3,\ldots,n-3$ and $j=i+2,i+3,\ldots,n-1$)}
In this case, we have considered $2\leq i\leq n-3$ and $i+2\leq j\leq n-1$.
From Case 2 as discussed above and Case 2 discussed in the proof of Lemma \ref{V 1 gen}, there are 
\[
\#\b=\left((q-3)-\tau_i^{(2)}\right)\cdot\sum_{j=i+2}^{n-1}\left((q-3)+\tau_j^{(2)}\right)
\] 
HF sequences such that given $a_i\neq b_i$ and $a_j\neq b_j$ for $j=i+2,i+3,\ldots,n-1$.
Hence for $i=2,3,\ldots,n-3$ and $j=i+2,i+3,\ldots,n-1$, the total number of HF sequences such that $a_i\neq b_i$ and $a_j\neq b_j$ is 
\[
\#\b=\sum_{i=2}^{n-3}\left(\left((q-3)-\tau_i^{(2)}\right)\cdot\sum_{j=i+2}^{n-1}\left((q-3)+\tau_j^{(2)}\right)\right) 
\]

Hence, for given an HF sequence $\a$, the total number of HF sequences $\b$ such that $d(\a,\b)=2$ and both the HF sequences are not differ at consecutive locations is the sum of sequences enumerated in Case 1, Case 2, Case 3 and Case 4, and the sum is  
    
    \begin{equation}
        \begin{split}       
        \#\b=& (q-2)^2+2\left[(q-2)\sum_{j=3}^{n-1}\left((q-3)+\tau_j^{(2)}\right)\right] \\
        &\hspace{0.2cm}+\sum_{i=2}^{n-3}\left(\left((q-3)-\tau_i^{(2)}\right)\cdot\sum_{j=i+2}^{n-1}\left((q-3)+\tau_j^{(2)}\right)\right) 
        \end{split}
        \label{eq: V 2 2nd part}
    \end{equation}
Hence, the HF sphere size $|H_2(\a)|$ follows from adding Equation (\ref{eq: V 2 1st part}) and Equation (\ref{eq: V 2 2nd part}).
And, it completes the proof.
    
\section{Sphere Size Table}
\label{App. Table}
We have listed HF sphere sizes $|H_r(\a)|$ for various HF sequences defined over alphabet $\mathcal{A}_4$ at Table \ref{HF Sphere Size Table}, Table \ref{HF Sphere Size for 4 length Table} and Table \ref{HF Sphere Size for 5 length Table} in this appendix. 
Recall that the size of the DNA alphabet $\{A,C,G,T\}$ is four.
Hence, the listed HF sphere sizes in these tables are also applicable to DNA sequences free from homopolymers. 
Therefore, in Table \ref{HF Sphere Size for 4 length Table} and Table \ref{HF Sphere Size for 5 length Table}, we have listed the HF sequences and HF sphere sizes over $\{A, C, G, T\}$. 
In Table \ref{HF Sphere Size Table}, we have listed sphere sizes $|H_r(\a)|$ for $\a=\a_{max}$ and $\a=\a_{min}$ of length $n=1,2,\ldots,10$ and radius $r=1,2,\ldots,n$. 
Recall that for $r=1,2$, the HF sequence $\a_{max}$ = $a_1a_2\ldots a_n$ with the condition $a_i$ = $a_{i+2}$ for $i=1,2,\ldots,n-2$, and the HF sequence $\a_{min}$ = $a_1a_2\ldots a_n$ with the condition $a_i$ = $a_{i+3}$ for $i=1,2,\ldots,n-3$.
In Table \ref{HF Sphere Size for 4 length Table}, we have classified HF DNA sequences, and therefore HF sequences over $\mathcal{A}_4$, on the bases of HF sphere size $|H_r(\a)|$, where $\a$ is HF DNA sequence of length $4$ and $r=1,2,3,4$. 
Similarly, in Table \ref{HF Sphere Size for 5 length Table}, we have classified HF DNA sequences on the bases of HF sphere size $|H_r(\a)|$, where $\a$ is an HF DNA sequence of length $5$ and $r=1,2,3,4,5$. 

\begin{table*}
\caption{The sphere size $|H_r(\a)|$ for HF sequences $\a_{min}$ and $\a_{max}$ of length $n$ over $\mathcal{A}_4$ with radius $r$ are illustrated.}
\begin{center}
\begin{tabular}{|c|m{4cm}|c|c|c|c|c|c|c|c|c|c|}
 \hline
 Length & Centre(s) &\multicolumn{10}{|c|}{Sphere Size for} \\  \cline{3-12}
$n$ &       &  $r=1$  &  $r=2$  & $r=3$ & $r=4$ & $r=5$ & $r=6$ & $r=7$ & $r=8$ & $r=9$ & $r=10$  \\ \hline
 1 & $\a_{max}=\a_{min}=a_1$;         &  3  &  -   & -  & -  & - & - & - & - & - & -  \\ \hline
 2 & $\a_{max}=\a_{min}=a_1a_2$;         &  4  &  7   & -  & -  & - & - & - & - & -  & - \\ \hline
 3 & $\a_{min}=a_1a_2a_3$; $a_1\neq a_3$ &  5  &  14  & 16 & -  & - & - & - & - & -   & -\\  \cline{2-12}
   & $\a_{max}=a_1a_2a_3$; $a_1=a_3$ &  6  &  12  & 17 & -  & - & - & - & - & -  & - \\   \hline
 4 & $\a_{min}=a_1a_2a_3a_4$; $a_1 = a_4$ &  6  &  21  & 44 & 36 & - & - & - & - & - & -   \\  \cline{2-12}
   % & $a_1a_2a_3a_4$; $a_1\neq a_3$ and $a_2\neq a_4$ &  6  &  22  & 42 & 37 & -& -  & - & - & - & -  \\   \cline{2-12}
   % & $a_1a_2a_3a_4$; Either ($a_1\neq a_3$ and $a_2 = a_4$)   &  7  &  22  & 39 & 39 & - & - & - & - & - & -  \\ 
   % & \hspace{1.35cm} or ($a_1 = a_3$ and $a_2 \neq a_4$)  &  &  &  &  &  &  &  &  &   & - \\ \cline{2-12}
   & $\a_{max}=a_1a_2a_3a_4$; \newline{$a_1 = a_3$ and $a_2 = a_4$} &  8  &  22  & 36 & 41 & - & - &  - & - & -& -   \\  \hline
 5 & $\a_{min}=a_1a_2a_3a_4a_5$; \newline{$a_1 = a_4$ and $a_2 = a_5$} & 7 & 29 & 79 & 127 & 81 & - & - & - & - & -  \\  \cline{2-12}
   % & $a_1a_2a_3a_4a_5$; Either ($a_1 = a_4$ and $a_2\neq a_5$) & 7 & 30 & 79 & 124 & 83 & - & - & -  & - & -  \\ 
   % & \hspace{1.7cm} or ($a_1\neq a_4$ and $a_2 = a_5$) &  &  &  &  &  &  &  &  &  & - \\ \cline{2-12}
   % & $a_1a_2a_3a_4a_5$; $a_1 = a_5$ and $a_2\neq a_4$ & 7 & 31 & 80 & 119 & 86 & -  & -& -  & - & -  \\ \cline{2-12}
   % & $a_1a_2a_3a_4a_5$; Either ($a_1 = a_3$ and $a_2 = a_5$) & 8 & 31 & 79 & 117 & 88 & - & - & -  & - & -  \\ 
   % & \hspace{1.7cm} or ($a_1 = a_4$ and $a_3 = a_5$) &  &  &  &  &  &  &  &  &   & - \\ \cline{2-12}
   % & $a_1a_2a_3a_4a_5$; $a_2 \neq a_4$ and either $a_1 = a_3 \neq a_5$ & 8 & 32 & 79 & 114 & 90 & - & - & -  & - & -  \\
   % & \hspace{1.7cm} or $a_1 \neq a_3 = a_5$ &  &  &  &  &  &  &  &  &  & - \\ \cline{2-12}
   % & $a_1a_2a_3a_4a_5$; $a_1 \neq a_3 \neq a_5$, $a_1\neq a_5$ and $a_2 = a_4$ & 8 & 33 & 76 & 117 & 89 & - & - & - & - & -  \\ \cline{2-12}
   % & $a_1a_2a_3a_4a_5$; $a_1 = a_5 \neq a_3$ and $a_2 = a_4$ & 8 & 33 & 77 & 115 & 90 & - & - & - & -  & - \\ \cline{2-12}
   % & $a_1a_2a_3a_4a_5$; $a_2 = a_4$ and either $a_1 = a_3 \neq a_5$ & 9 & 34 & 76 & 110 & 94 & - & - & -  & -& -  \\ 
   % & \hspace{1.7cm} or $a_1 \neq a_3 = a_5$ &  &  &  &  &  &  &  &  &  & - \\ \cline{2-12}
   % & $a_1a_2a_3a_4a_5$; $a_1 = a_3 = a_5$ and $a_2 \neq a_4$ & 9 & 34 & 77 & 108 & 95 & - & - & - & -  & - \\ \cline{2-12}
   & $\a_{max}=a_1a_2a_3a_4a_5$; \newline{$a_1 = a_3 = a_5$ and $a_2 = a_4$} & 10 & 36 & 74 & 104 & 99 & -  & -& - & - & -  \\   \hline
 6 & $\a_{min}=a_1a_2\ldots a_6$; \newline{$a_i = a_{i+3}$ for $i=1,2,3$} & 8 & 38 & 124 & 269 & 350 & 182 & - & - & - & -  \\  \cline{2-12}
   & $\a_{max}=a_1a_2\ldots a_6$; \newline{$a_i = a_{i+2}$ for $i=1,2,3,4$} & 12 & 54 & 136 & 238 & 292 & 239 & - & - & - & -  \\  \hline
 7 & $\a_{min}=a_1a_2\ldots a_7$; \newline{$a_i = a_{i+3}$ for $i=1,2,3,4$}      & 9  & 48  & 180 & 476 & 861  & 932  & 409  & -& - & -  \\  \cline{2-12}
   & $\a_{max}=a_1a_2\ldots a_7$; \newline{$a_i = a_{i+2}$ for $i=1,2,3,4,5$}    & 14 & 76  & 230 & 480 & 734  & 804  & 577 & - & -  & -  \\  \hline
 8 & $\a_{min}=a_1a_2\ldots a_8$; \newline{$a_i = a_{i+3}$ for $i=1,2,3,4,5$}      & 10 & 59  & 248 & 761 & 1702 & 2626 & 2422 & 919 & -  & -  \\  \cline{2-12}
   & $\a_{max}=a_1a_2\ldots a_8$; \newline{$a_i = a_{i+2}$ for $i=1,2,\ldots,6$} & 16 & 102 & 364 & 886 & 1608 & 2198 & 2180 & 1393 & -  & -  \\  \hline
 9 & $\a_{min}=a_1a_2\ldots a_9$; \newline{$a_i = a_{i+3}$ for $i=1,2,\ldots,6$}      & 11 & 71 & 329 & 1138 & 2977 & 5758 & 7715 & 6179 & 2065 & -   \\  \cline{2-12}
   & $\a_{max}=a_1a_2\ldots a_9$; \newline{$a_i = a_{i+2}$ for $i=1,2,\ldots,7$} & 18 & 132 & 546 & 1528 & 3202 & 5180 & 6434 & 5840 & 3363  & -  \\  \hline
 10& $\a_{min}=a_1a_2\ldots a_{10}$; \newline{$a_i = a_{i+3}$ for $i=1,2,\ldots,7$}  & 12 & 84 & 424 & 1622 & 4806 & 10963 & 18638 & 22002 & 15540  & 4640 \\ \cline{2-12}
   & $\a_{max}=a_1a_2\ldots a_{10}$; \newline{$a_i = a_{i+2}$ for $i=1,2,\ldots,8$} & 20 & 166 & 784 & 2494 & 5932 & 11030 & 16200 & 18494 & 15492  & 8119 \\  \hline
\end{tabular}
\end{center}
\label{HF Sphere Size Table}
\end{table*}

\begin{table*}
\caption{The HF sphere size for DNA sequences $\a$ of length $4$ are illustrated with radius $r$.}
\begin{center}
\begin{tabular}{|c|cccc|c|c|c|c|}
 \hline
 Sr. No. & \multicolumn{4}{|c|}{Centre} &\multicolumn{4}{|c|}{Sphere Size for} \\  \cline{6-9}
 &   \multicolumn{4}{|c|}{$\a$}              & $r=1$ & $r=2$ & $r=3$ & $r=4$   \\ \hline 									
1	    &	ACGA	&	CAGC	&	GACG	&	TACT	&6	&21	&44	& 36 \\
	&	ACTA	&	CATC	&	GATG	&	TAGT	&&&& \\
	&	AGCA	&	CGAC	&	GCAG	&	TCAT	&&&& \\
	&	AGTA	&	CGTC	&	GCTG	&	TCGT	&&&& \\
	&	ATCA	&	CTAC	&	GTAG	&	TGAT	&&&& \\
	&	ATGA	&	CTGC	&	GTCG	&	TGCT	&&&& \\ \hline 
2	    &	ACGT	&	CAGT	&	GACT	&	TACG	&6	&22	&42	&37 \\
	&	ACTG	&	CATG	&	GATC	&	TAGC	&&&& \\
	&	AGCT	&	CGAT	&	GCAT	&	TCAG	&&&& \\
	&	AGTC	&	CGTA	&	GCTA	&	TCGA	&&&& \\
	&	ATCG	&	CTAG	&	GTAC	&	TGAC	&&&& \\
	&	ATGC	&	CTGA	&	GTCA	&	TGCA	&&&& \\ \hline 
3	    &	ACAG	&	CACG	&	GACA	&	TACA	&7	&22	&39	&39 \\
	&	ACAT	&	CACT	&	GAGC	&	TAGA	&&&& \\
	&	ACGC	&	CAGA	&	GAGT	&	TATC	&&&& \\
	&	ACTC	&	CATA	&	GATA	&	TATG	&&&& \\
	&	AGAC	&	CGAG	&	GCAC	&	TCAC	&&&& \\
	&	AGAT	&	CGCA	&	GCGA	&	TCGC	&&&& \\
	&	AGCG	&	CGCT	&	GCGT	&	TCTA	&&&& \\
	&	AGTG	&	CGTG	&	GCTC	&	TCTG	&&&& \\
	&	ATAC	&	CTAT	&	GTAT	&	TGAG	&&&& \\
	&	ATAG	&	CTCA	&	GTCT	&	TGCG	&&&& \\
	&	ATCT	&	CTCG	&	GTGA	&	TGTA	&&&& \\
	&	ATGT	&	CTGT	&	GTGC	&	TGTC	&&&& \\ \hline 
4       &	ACAC	&	CACA	&	GAGA	&	TATA	&8	&22	&36	&41 \\
	&	AGAG	&	CGCG	&	GCGC	&	TCTC	&&&& \\
	&	ATAT	&	CTCT	&	GTGT	&	TGTG	&&&& \\  \hline 									
\end{tabular}
\end{center}
\label{HF Sphere Size for 4 length Table}
\end{table*}

\begin{table*}
\caption{The HF sphere size for DNA sequences $\a$ of length $5$ are illustrated with radius $r$.}
\begin{center}
\begin{tabular}{|c|cccc|c|c|c|c|c|}
 \hline
 Sr. No. & \multicolumn{4}{|c|}{Centre} &\multicolumn{5}{|c|}{Sphere Size for} \\  \cline{6-10}
 &   \multicolumn{4}{|c|}{$\a$}              & $r=1$ & $r=2$ & $r=3$ & $r=4$ & $r=5$  \\ \hline 
1 &	ACGAC	&	CAGCA	&	GACGA	&	TACTA	&7	&29	&79	&127	&81	\\
&	ACTAC	&	CATCA	&	GATGA	&	TAGTA	&&&&&	\\
&	AGCAG	&	CGACG	&	GCAGC	&	TCATC	&&&&&	\\
&	AGTAG	&	CGTCG	&	GCTGC	&	TCGTC	&&&&&	\\
&	ATCAT	&	CTACT	&	GTAGT	&	TGATG	&&&&&	\\
&	ATGAT	&	CTGCT	&	GTCGT	&	TGCTG	&&&&&	\\ \hline 
2&	ACGAT	&	CAGCT	&	GACGT	&	TACGA	&7	&30	&79	&124&	83	\\
&	ACGTC	&	CAGTA	&	GACTA	&	TACTG	&&&&&	\\
&	ACTAG	&	CATCG	&	GATCA	&	TAGCA	&&&&&	\\
&	ACTGC	&	CATGA	&	GATGC	&	TAGTC	&&&&&	\\
&	AGCAT	&	CGACT	&	GCAGT	&	TCAGC	&&&&&	\\
&	AGCTG	&	CGATG	&	GCATC	&	TCATG	&&&&&	\\
&	AGTAC	&	CGTAG	&	GCTAC	&	TCGAC	&&&&&	\\
&	AGTCG	&	CGTCA	&	GCTGA	&	TCGTA	&&&&&	\\
&	ATCAG	&	CTACG	&	GTACT	&	TGACG	&&&&&	\\
&	ATCGT	&	CTAGT	&	GTAGC	&	TGATC	&&&&&	\\
&	ATGAC	&	CTGAT	&	GTCAT	&	TGCAG	&&&&&	\\
&	ATGCT	&	CTGCA	&	GTCGA	&	TGCTA	&&&&&	\\ \hline 
3&	ACGTA	&	CAGTC	&	GACTG	&	TACGT	&7	&31	&80	&119	&86	\\
&	ACTGA	&	CATGC	&	GATCG	&	TAGCT	&&&&&	\\
&	AGCTA	&	CGATC	&	GCATG	&	TCAGT	&&&&&	\\
&	AGTCA	&	CGTAC	&	GCTAG	&	TCGAT	&&&&&	\\
&	ATCGA	&	CTAGC	&	GTACG	&	TGACT	&&&&&	\\
&	ATGCA	&	CTGAC	&	GTCAG	&	TGCAT	&&&&&	\\ \hline 
4&	ACAGC	&	CACGA	&	GACGC	&	TACTC	&8	&31	&79	&117	&88	\\
&	ACATC	&	CACTA	&	GAGCA	&	TAGTG	&&&&&	\\
&	ACGAG	&	CAGCG	&	GAGTA	&	TATCA	&&&&&	\\
&	ACTAT	&	CATCT	&	GATGT	&	TATGA	&&&&&	\\
&	AGACG	&	CGACA	&	GCAGA	&	TCATA	&&&&&	\\
&	AGATG	&	CGCAG	&	GCGAC	&	TCGTG	&&&&&	\\
&	AGCAC	&	CGCTG	&	GCGTC	&	TCTAC	&&&&&	\\
&	AGTAT	&	CGTCT	&	GCTGT	&	TCTGC	&&&&&	\\
&	ATACT	&	CTACA	&	GTAGA	&	TGATA	&&&&&	\\
&	ATAGT	&	CTCAT	&	GTCGC	&	TGCTC	&&&&&	\\
&	ATCAC	&	CTCGT	&	GTGAT	&	TGTAG	&&&&&	\\
&	ATGAG	&	CTGCG	&	GTGCT	&	TGTCG	&&&&&	\\ \hline 
5&	ACAGT	&	CACGT	&	GACTC	&	TACGC	&8	&32	&79	&114	&90	\\
&	ACATG	&	CACTG	&	GAGCT	&	TAGCG	&&&&&	\\
&	ACGTG	&	CAGTG	&	GAGTC	&	TATCG	&&&&&	\\
&	ACTGT	&	CATGT	&	GATCT	&	TATGC	&&&&&	\\
&	AGACT	&	CGATA	&	GCATA	&	TCAGA	&&&&&	\\
&	AGATC	&	CGCAT	&	GCGAT	&	TCGAG	&&&&&	\\
&	AGCTC	&	CGCTA	&	GCGTA	&	TCTAG	&&&&&	\\
&	AGTCT	&	CGTAT	&	GCTAT	&	TCTGA	&&&&&	\\
&	ATACG	&	CTAGA	&	GTACA	&	TGACA	&&&&&	\\
&	ATAGC	&	CTCAG	&	GTCAC	&	TGCAC	&&&&&	\\
&	ATCGC	&	CTCGA	&	GTGAC	&	TGTAC	&&&&&	\\
&	ATGCG	&	CTGAG	&	GTGCA	&	TGTCA	&&&&&	\\ \hline 
6&	ACGCT	&	CAGAT	&	GACAT	&	TACAG	&8	&33	&76	&117	&89	\\
&	ACTCG	&	CATAG	&	GATAC	&	TAGAC	&&&&&	\\
&	AGCGT	&	CGAGT	&	GCACT	&	TCACG	&&&&&	\\
&	AGTGC	&	CGTGA	&	GCTCA	&	TCGCA	&&&&&	\\
&	ATCTG	&	CTATG	&	GTATC	&	TGAGC	&&&&&	\\
&	ATGTC	&	CTGTA	&	GTCTA	&	TGCGA	&&&&&	\\ \hline 
7&	ACGCA	&	CAGAC	&	GACAG	&	TACAT	&8	&33	&77	&115	&90	\\
&	ACTCA	&	CATAC	&	GATAG	&	TAGAT	&&&&&	\\
&	AGCGA	&	CGAGC	&	GCACG	&	TCACT	&&&&&	\\
&	AGTGA	&	CGTGC	&	GCTCG	&	TCGCT	&&&&&	\\
&	ATCTA	&	CTATC	&	GTATG	&	TGAGT	&&&&&	\\
&	ATGTA	&	CTGTC	&	GTCTG	&	TGCGT	&&&&&	\\ \hline 
\multicolumn{10}{r}{{Table \ref{HF Sphere Size for 5 length Table} – continued on next page}}  \\ 
\end{tabular}
\end{center}
\label{HF Sphere Size for 5 length Table}
\end{table*}

\begin{table*}
\begin{center}
\begin{tabular}{|c|cccc|c|c|c|c|c|}
\multicolumn{10}{l}{{Table \ref{HF Sphere Size for 5 length Table} – continued from previous page}} \\ \hline
 Sr. No. & \multicolumn{4}{|c|}{Centre} &\multicolumn{5}{|c|}{Sphere Size for} \\  \cline{6-10}
 &   \multicolumn{4}{|c|}{$\a$}              & $r=1$ & $r=2$ & $r=3$ & $r=4$ & $r=5$  \\ \hline 
8&	ACACG	&	CACAG	&	GACAC	&	TACAC	&9	&34	&76	&110	&94	\\
&	ACACT	&	CACAT	&	GAGAC	&	TAGAG	&&&&&	\\
&	ACGCG	&	CAGAG	&	GAGAT	&	TATAC	&&&&&	\\
&	ACTCT	&	CATAT	&	GATAT	&	TATAG	&&&&&	\\
&	AGAGC	&	CGAGA	&	GCACA	&	TCACA	&&&&&	\\
&	AGAGT	&	CGCGA	&	GCGCA	&	TCGCG	&&&&&	\\
&	AGCGC	&	CGCGT	&	GCGCT	&	TCTCA	&&&&&	\\
&	AGTGT	&	CGTGT	&	GCTCT	&	TCTCG	&&&&&	\\
&	ATATC	&	CTATA	&	GTATA	&	TGAGA	&&&&&	\\
&	ATATG	&	CTCTA	&	GTCTC	&	TGCGC	&&&&&	\\
&	ATCTC	&	CTCTG	&	GTGTA	&	TGTGA	&&&&&	\\
&	ATGTG	&	CTGTG	&	GTGTC	&	TGTGC	&&&&&	\\ \hline 
9&	ACAGA	&	CACGC	&	GAGCG	&	TATCT	&9	&34	&77	&108	&95	\\
&	ACATA	&	CACTC	&	GAGTG	&	TATGT	&&&&&	\\
&	AGACA	&	CGCAC	&	GCGAG	&	TCTAT	&&&&&	\\
&	AGATA	&	CGCTC	&	GCGTG	&	TCTGT	&&&&&	\\
&	ATACA	&	CTCAC	&	GTGAG	&	TGTAT	&&&&&	\\
&	ATAGA	&	CTCGC	&	GTGCG	&	TGTCT	&&&&&	\\ \hline 
10&	ACACA	&	CACAC	&	GAGAG	&	TATAT	&10	&36	&74	&104	&99	\\
&	AGAGA	&	CGCGC	&	GCGCG	&	TCTCT	&&&&&	\\
&	ATATA	&	CTCTC	&	GTGTG	&	TGTGT	&&&&&	\\ \hline 
\end{tabular}
\end{center}
\end{table*}

\end{document}